\documentclass[%
 reprint,
superscriptaddress,
 amsmath,amssymb,
 aps,
 pra,
]{revtex4-2}
\DeclareMathAlphabet{\pazocal}{OMS}{zplm}{m}{n}

\usepackage{graphicx}
\usepackage{dcolumn}
\usepackage{bm}
\usepackage{xcolor}
\usepackage{braket}
\usepackage[colorlinks=true,linkcolor=blue,citecolor=blue] {hyperref}

\begin{document}

\preprint{APS/123-QED}

\title{Topological argument for robustness of coherent states in quantum optics}

\author{Saumya Biswas}
\affiliation{%
Department of Mechanical Engineering,
University of Maryland, College Park, MD 20742, USA
}%
\affiliation{Institute for Physical Science and Technology,
University of Maryland, College Park, MD 20742, USA}

\author{Amrit De}
\affiliation{
Apsidal, 3600 Lime Street, Riverside, CA 92501, USA\\
}%

\author{Avik Dutt}%
\email{avikdutt@umd.edu}
\affiliation{%
Department of Mechanical Engineering,
University of Maryland, College Park, MD 20742, USA
}%
\affiliation{Institute for Physical Science and Technology,
University of Maryland, College Park, MD 20742, USA}
\affiliation{National Quantum Laboratory (QLab) at Maryland, College Park, MD 20742, USA}


\date{\today}




\begin{abstract}
Coherent states, being the closest analog to classical states of wave systems, are well known to possess special properties that set them apart from most other quantum optical states. For example, they are robust against photon loss and do not easily get entangled upon interaction with a beamsplitter, and hence are called ``pointer states'', which is often attributed to them being eigenstates of the annihilation operator. Here we provide insights into a topological argument for their robustness using two separate but exact mappings of a prototypical quantum optics model - the driven Jaynes-Cummings model. The first mapping is based on bosonization and refermionization of the Jaynes-Cummings model into the fermionic Su-Schrieffer-Heeger model hosting zero-energy topologically protected edge states. The second mapping is based on the algebra of deformed f-oscillators. We choose these mappings to explicitly preserve the translational symmetry of the model along a Fock-state ladder basis, which is important for maintaining the symmetry-protected topology of such 1D lattices. In addition, we show that the edge state form is preserved even when certain chiral symmetry is broken, corresponding to a single-photon drive for the quantum optics model that preserves the coherent state; however, the addition of two-photon drive immediately disturbs the edge state form, as confirmed by numerical simulations of the mapped SSH model; this is expected since two-photon drive strongly perturbs the coherent state into a squeezed state. Our theory sheds light on a fundamental reason for the robustness of coherent states, both in existence and entanglement -- an underlying connection to topology.

\end{abstract}
\maketitle

\section{Introduction}
Coherent States (CS) (\cite{glauber1963quantum,glauber1963photon,glauber1963coherent,sudarshan1963equivalence}) are the closest quantum mechanical analogues of classical states and are especially important in the theories of quantum information and quantum measurement \cite{haroche2006exploring}. Substantial recent research has explored how to use photonic structures endowed with nontrivial topology \cite{price_roadmap_2022, ozawa_topological_2019} for the robust protection of certain properties of optical states, including classical coherent states, nonclassical states, partially coherent light, and two-photon quantum correlations \cite{barik_topological_2018, blanco-redondo_topological_2018, wang_topologically_2019, wang_topological_2019, dai_topologically_2022}. 

However, even in the absence of external photonic structures, coherent states are robust in themselves, and hence the question arises if there is a nontrivial topological explanation for the robustness exhibited by coherent states. As examples of robustness, coherent states show intriguing properties in terms of pure state evolution when subjected to interactions with the environment, and they do not generate entanglement when incident on a beam splitter \cite[chapter-3,4]{haroche2006exploring}. Such immunity to entanglement is a property also shown by topological edge states (ES), as long as the global topology of the manifold is preserved and the energy bandgap is not closed. Other recent works have investigated edge states \cite{dangel2018topological,lieu2020tenfold} as edge dark states due to the same immunity to environmental entanglement \cite{yang2023symmetry,diehl2011topology}, with emphasis on the stability of topologically nontrivial states against different channels of dissipation \cite{leefmans_topological_2022, sridhar_quantized_2024-2}.

Here we address the question of robustness exhibited by coherent states and provide topological arguments for the origin of such robustness. We achieve this by exactly mapping driven-dissipative Hamiltonian models that facilitate an understanding of coherent states as topologically protected edge states. The topological model emphasises the role of underlying symmetries (or antisymmetries) for the stability of edge states \cite{diehl2011topology} that represents the coherent states. 
%
A less restrictive requirement of exact pointer states even if symmetries are broken is sufficient for immunity from environmental entanglement \cite[chapter-4.4.5]{haroche2006exploring}.
In fact, CS are exact pointer states only (and not dark states) when subject to photon loss channels. 
%
It is straightforward to meet the requirements of pointer states in the mapped model.  
%
We show that different coherent interactions, incoherent interactions (e.g. an environmental loss channel) and photon drives can be mapped to the translationally symmetric fermionic lattice, the symmetry of which governs the existence and stability of edge states.

We investigate a driven-dissipative Jaynes-Cummings quantum-optics model \cite{kastoryano2019topological, fischer_pulsed_2018} in this paper. 
We show that this system can be injectively mapped onto a fermionic Su–Schrieffer–Heeger (fSSH) model that preserves translational symmetry, and is known to support edge states. 
In a parallel approach , we also show a translationally symmetric map to bosonic models with nonlinear atom-photon interactions  -- the so-called  
 nonlinear Jaynes Cummings (NLJC) Hamiltonian \cite{miry2012generation}. 
Although their Hilbert spaces are different, the two models have the same matrices in their respective representations. 
Note that this theory is only applicable for the restricted case of the coherent state being an eigenstate of the nondissipative part of the Hamiltonian.


\begin{figure}[h]
\includegraphics[width=3in]{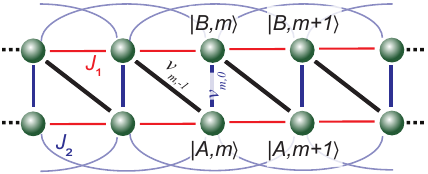}
\caption{\label{fig:lattice} Schematic of the 1-D lattice Hamiltonian $H_c$ (Eq. \eqref{eq_Rudner_Ham}) with internal states $|A\rangle$ and $|B\rangle$ with an energy difference $\Delta$. The $v_{m,j}$ terms couple sites $|m\rangle$ with different internal states, while the $J_j$ terms couple sites of the same internal state ($H^{si}_c$ in Eq. \eqref{eq_site_ind}). For $J_1=J_2=0$, $v_{m,j}$ independent of $m$, this reduces to the translationally invariant SSH model. 
}
\end{figure}


\section{Jaynes-Cummings Model in the Fock Space Lattice}
We first review a one-dimensional two-band lattice model and its connection to the Jaynes-Cummings model in the photon number basis. Consider a Hamiltonian for a 1D ladder type lattice: 
\begin{equation}
H_c=\Delta\sum_{m}  |B,m \rangle \langle B, m| + \sum\limits_{m,\langle j\rangle} v_{m,j} |B,m+j \rangle \langle A,m| + \textrm{H.c.} 
 \label{eq_Rudner_Ham}
\end{equation}
where sites $m\in \mathbb{Z}$ and each site $m$ consists of the two internal states $|A,m\rangle$ and $|B,m\rangle$,
$\langle j\rangle$ denotes the sum over nearest neighbor sites (see Fig. \ref{fig:lattice}) and $\Delta$ is the difference in the energies associated with  $|B\rangle$ and $|A\rangle$. 
Note that the inter-site hopping term in Eq.-\eqref{eq_Rudner_Ham} only couples $|B,m\rangle$ to $|A,m+1\rangle$. 
In addition, one can introduce nearest and next-nearest neighbor hoppings, $J_1$ and $J_2$ respectively, that preserve the internal state:
\begin{equation}
 H^{si}_c = \sum_{j>0} \ \sum_{S\in{A,B}} J_j | S,m+j\rangle \langle S,m | + H.c.\label{eq_site_ind}
\end{equation}

\begin{figure}[h]
\includegraphics[width=3.2in]{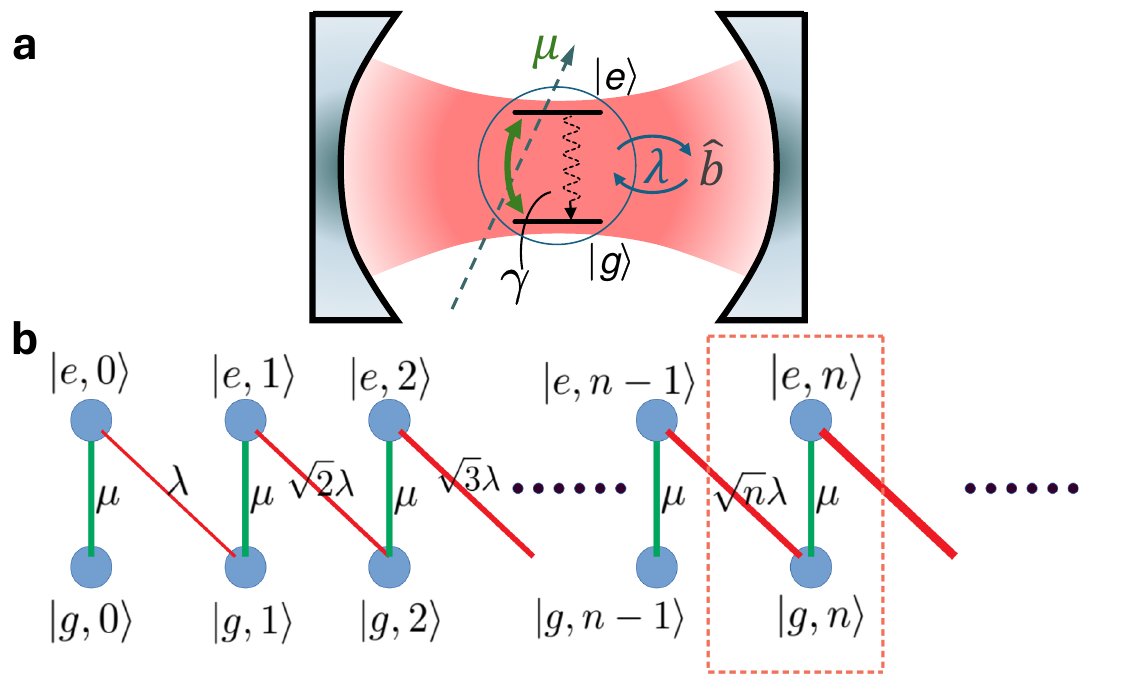}
\caption{\label{fig:FSL} (a) An atom of transition frequency $\omega_a$ is driven resonantly by the cavity mode b (with oscillation frequency $\omega_c$) and a classical drive of frequency $\omega_L$. $\lambda$ and $\mu$ denote the strength of atom-cavity interaction and atom-classical drives respectively. $|e\rangle$ and $|g\rangle$ denote the atom's excited and ground states. The excited state $\ket{e}$ exhibits a spontaneous decay at a rate $\gamma$. (b) A Fock Space Lattice (FSL) representation of the single-bosonic-mode Jaynes-Cummings Hamiltonian of the setup in (a). 
The green and red lines denote the classical resonant drive at $\omega_c$ and the atom drive and the (quantum cavity drive) on the atom (as in (a)). The red edges between Fock states $|n\rangle$ with a photon number difference of 1 get stronger $\propto \sqrt{n}$. This FSL representation maps to Fig. \ref{fig:lattice} and Eq. \eqref{eq_Rudner_Ham} with $J_1=J_2 = 0$ and $v_{m,0}=\mu, v_{m,-1}=\lambda\sqrt{m}$.}
\end{figure}

The starting point of our analysis is a map of the Jaynes-Cummings model to the Hamiltonian in Eq. \eqref{eq_Rudner_Ham} where the lattice-site index comes from the photon number basis $|m\rangle$ -- an approach that has recently been dubbed the Fock-Space-Lattice (FSL) picture \cite{saugmann2023fock, cai2021topological}. Specifically, consider a single quantized cavity mode with annihilation operator $b$ coupled with strength $\lambda$ to a two-level atom. The atomic state has raising and lowering operators $\sigma^{\pm}$ and is subjected to an external classical drive of strength $\mu$ \cite{kastoryano2019topological, solano_strong-driving-assisted_2003, zou_scheme_2004, bermudez2012robust}. 
%
In this paper we only consider the case where the classical drive ($\omega_L$) and the cavity mode ($\omega_c$) are resonant with the atomic transition ($\omega_a$), i.e., $\omega_L = \omega_a = \omega_c$.
The simplified Hamiltonian is
\begin{equation}
\mathcal{H}^d_{JC} = \left(\lambda \sigma^+ b + \lambda \sigma^- b^{\dagger} + \mu \sigma^+ + \mu \sigma^- \right) \label{eq_dJC_min}.
\end{equation}
Now, this model maps to Eq. \eqref{eq_Rudner_Ham} with a photon-number ($m$)-dependent coupling $v_{m,-1} \propto \lambda \sqrt{m}$ that \textit{breaks translational symmetry} under the usual FSL approach \cite{saugmann2023fock, cai2021topological} which expands $b$ as 
\begin{equation}
b = \sum\limits_{n=0}^{\infty} \sqrt{n+1} | n \rangle \langle n +1 | \ \ \ \ \ \label{eq_b}
\end{equation}
This is depicted in Fig. \ref{fig:FSL}). 
The $\sqrt{n+1}$ term appears in the hopping amplitudes for all known mappings from the JC model to the SSH model \cite{saugmann2023fock,cai2021topological}. 

In this paper, we instead propose two maps which preserve the translational symmetry, and relate the robustness of coherent states to topological protection. The first approach maps the Jaynes-Cummings model to the fermionic SSH model using a bosonization and partial refermionization approach. The second approach uses deformed f-oscillators.

Before proceeding to these two maps, we briefly discuss the single-cavity-mode FSL approach that breaks translation invariance, whereby a coherent state emerges as a dark state \cite{kastoryano2019topological} when spontaneous decay of the atom is introduced into $\mathcal{H}^d_{JC}$ (Eq. \eqref{eq_dJC_min}). This corresponds to an on-site decay on the $|B\rangle$ modes in $H_c$ (Eq. \eqref{eq_Rudner_Ham}). Note that such a single-cavity-mode model is different from the multimode or multi-atom extensions of the Jaynes-Cummings models that are frequently studied for constructing quantum topology in FSLs \cite{cai2021topological, deng_observing_2022, saugmann2023fock}).
Eq. \eqref{eq_dJC_min} can be mapped to Eq. \eqref{eq_Rudner_Ham} with $v_{m,0}=\mu$ and $v_{m,-1}=\lambda\sqrt{m}$. The $\sqrt{m}$ in $v_{m,-1}$ precludes translational symmetry and an exact mapping to a SSH model. In the absence of a photon loss channel but in the presence of atom decay characterized by the jump operator $L_m=\sqrt{\gamma} |A, m \rangle \langle B, m |$, a dark stationary state localized on the $A$ sublattice is realized of the form,
\begin{eqnarray}
| \bar{\psi} \rangle &=& \sum\limits_{m=0}^\infty \bar{\phi}_m | A, m \rangle, {\rm thus},\  L_m | \bar{\psi} \rangle = 0 \ \ \label{eq_L_z}\\
| \bar{\psi} \rangle &=& e^{-\left(-\mu/\lambda \right)^2/2}\sum\limits_{m=0}^{\infty} \frac{\left(-\mu/\lambda \right)^m}{\sqrt{m!}} | A, m \rangle \label{eq_dark_state_dJC} 
\end{eqnarray}
The bosonic coherent state $|\alpha=-\mu/\lambda\rangle$ represented by Eq. \eqref{eq_dark_state_dJC} is an eigenstate of $\mathcal{H}^d_{JC}$ with energy $E=0$, as can be verified by direct substitution into Eq. \eqref{eq_dJC_min}.

\section{Map to the Fermionic SSH model and fermion-boson duality}
Next, we discuss the first mapping onto the fSSH model, characterized by $H_{SSH}$.
This can be recast in the form of Eq. \eqref{eq_Rudner_Ham}, by identifying $v_{m,-1} = t_{\text{intercell}} = w$, $v_{m,0} = t_{\text{intracell}} = v$, and $v_{m,i} = 0$ for all $i \neq 0, -1$. For a semi-infinite lattice truncated at the left end, the left edge state (ES) is a zero-energy eigenstate $|\ell \rangle$ of $H_{SSH}$ \cite{tzortzakakis2022topological}. This ES has support exclusively on the A sublattice:
\begin{eqnarray}
|\ell\rangle= \sum_{m} \phi_m |A,m\rangle = c^{\ell}_0 \sum\limits_{m=0}^{\infty} (-v/w)^{m} |A,m\rangle  \label{eq_left_edge_state}
\end{eqnarray}
where we have used the normalization $|c^{\ell}_0|^{2}=1-(v/w)^2 $.
The state $|\ell \rangle$ in Eq. \eqref{eq_left_edge_state} is normalizable only in the topologically nontrivial regime ($w > v$), which features an ES with $E = 0$. However, both Eq. \eqref{eq_dark_state_dJC} and Eq. \eqref{eq_left_edge_state} result from the recurrence relation
\begin{eqnarray}
v_{m,0} \phi_m + v_{m,-1} \phi_{m+1} = 0,
\end{eqnarray}
as derived by substituting the ansatz from Eq. \eqref{eq_L_z} into Eq. \eqref{eq_dJC_min} and \eqref{eq_Rudner_Ham} respectively. The zero-energy solution (Eq. \eqref{eq_dark_state_dJC}) to this recurrence relation exists for both $\mu > \lambda$ and $\mu < \lambda$. This is because, unlike a geometric series (as in the fSSH case), the region of convergence for the exponential function is infinite. This ensures that the normalization of $\langle\psi|\psi\rangle$ is finite and can be normalized to 1. The inclusion of $\sqrt{m}$ in $v_{m,-1}$ helps achieve convergence in both regimes.


\begin{figure}[h]
\includegraphics[height=5.4cm]{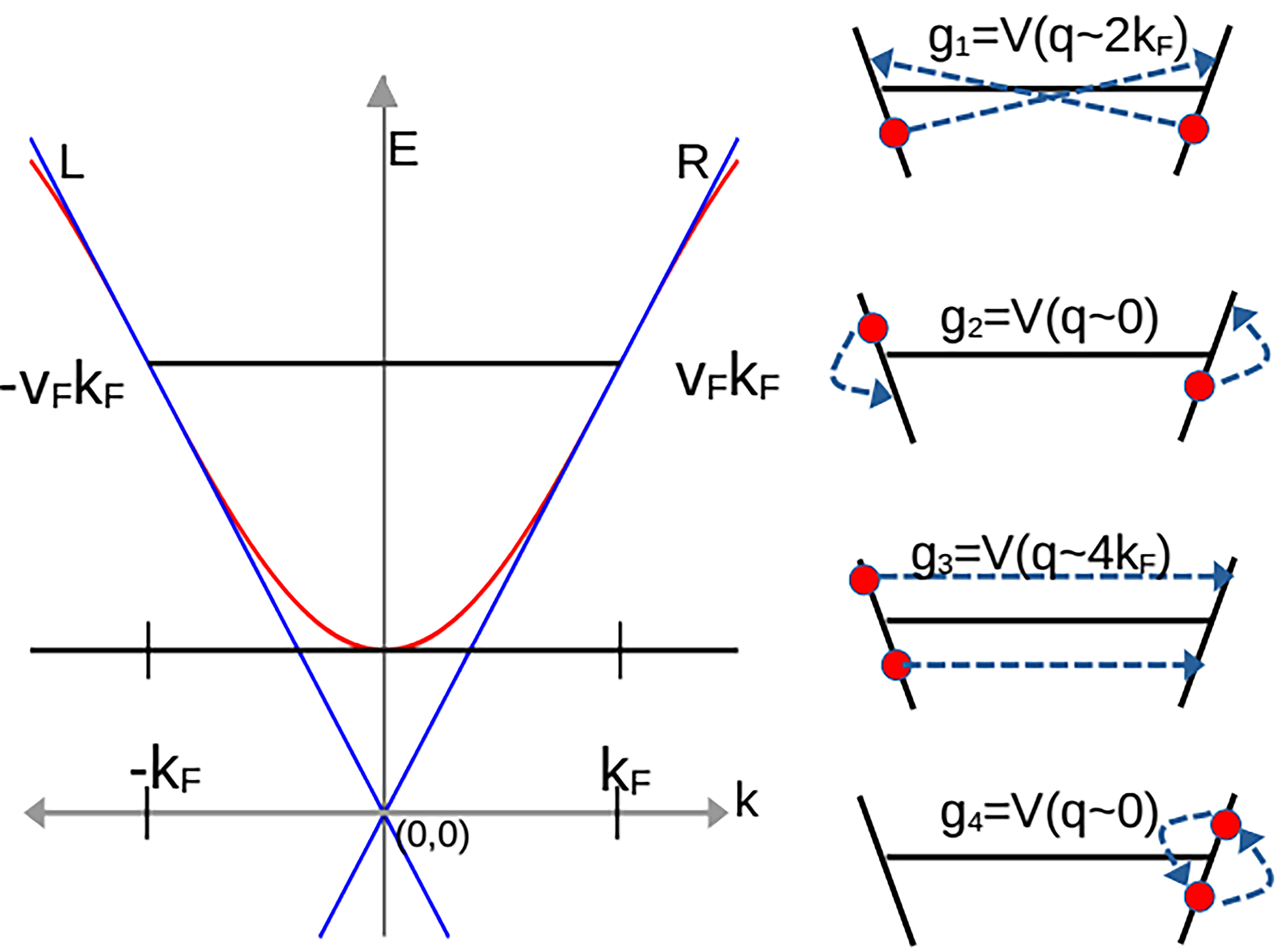}
\caption{\label{fig:disp} Linearized dispersion and the filled infinite Dirac sea in bosonization theory of 1D fermions. $k_F$($v_F$) is the Fermi wave-vector (velocity) around which the dispersion is linearized. The wavevector $q$-dependent interactions $V(q)$ are classified into four kinds $g_1$ through $g_4$. $g_2$ and $g_4$ do not change particle's branch: L (left) or R (right) and little $q$ is involved. $g_1$ and $g_3$ involve substantial momentum transfer as well transfer of branch. }
\end{figure}


We now propose a hitherto unused mapping idea from the bosonic Jaynes-Cummings model to the fSSH model. An important example of fermion-boson duality in 1D is provided by the coherent (collective) fermionic excitations of interacting 1D fermions, which can be bosonized as, 
\begin{eqnarray}
b^{\dagger}_q = \left(\frac{2\pi}{L|q|} \right)^{\frac{1}{2}}\sum\limits_{r=+1,-1} \Theta(rq) \rho_r^{\dagger}(q)  \label{eq_bq_dag}
\end{eqnarray}
where $:\rho_r^{\dagger}(q):= \sum\limits_k c^{\dagger}_{r,k+q} c_{r,k} $ is the fermion density operator,  on the left L ($r$=-1) and right R ($r$=1) branches (Fig. \ref{fig:disp}) respectively \cite{giamarchi2003quantum}. :: indicates normal ordering. $\Theta(\cdot)$ is the Heaviside step function. For a specific choice of $q$ (Eq. \eqref{eq_bq_dag} is defined for $q \neq 0$ only), only one density operator (either left or right) survives in Eq. \eqref{eq_bq_dag}. Thus, their linear combination can be taken to define a single annihilation operator, 
$ 
c(q) = \Theta(q) c_R + \Theta(-q) c_L \label{eq:c_lin}
$. 
We denote the prefactor by $K(q)=\sqrt{2\pi/L|q|}$ to arrive at,
\begin{equation}
b^{\dagger}_q = K\sum\limits_{k=-\infty}^{k_F} c^{\dagger}_{k+q} c_k \label{eq_b_sub}
\end{equation}
This operator raises any given $k$-state in the first Brillouin zone to a value higher by $q$, which is a $k$-vector of the reciprocal lattice. Since this expansion of the boson operator in Eq. \eqref{eq_b_sub} has no site-index dependence (such as $\sqrt{k}$ or a different $k$-dependent factor (unlike Eq. \eqref{eq_b}), we can aim to approach a map to a translationally symmetric model (e.g. the fSSH model). Using Eq. \eqref{eq_b_sub} in Eq. \eqref{eq_dJC_min} and choosing $q=1$ in suitable units, we obtain,
\begin{eqnarray}
\mathcal{H}^{(1)}_{JC} &=& t_{\rm inter} \left(\sigma^+ \sum_k c^{\dagger}_{k} c_{k+1} + \sigma^- \sum_k c^{\dagger}_{k+1} c_k \right) \nonumber\\
& &+ t_{\rm intra} \left( \sigma^+ + \sigma^- \right)  \label{eq_dJC_RM}
\end{eqnarray}
with $t_{\rm inter}=\lambda K$, $t_{\rm intra}=\mu$. Thus, the JC model can be mapped to a fSSH model. Manifestly, the $t_{\rm inter}$ term couples adjacent sites with opposite internal states ($\ket{A}$,  $\ket{B}$ in Fig. \ref{fig:lattice} and Eq. \eqref{eq_Rudner_Ham}'s $v_{m,-1}$), as denoted by the simultaneous action of $\sigma_{+}$ and hopping from site $k+1$ to $k$ via $c^{\dagger}_{k} c_{k+1}$, or the reverse. The $t_{\rm intra}$ term couples internal states on the same site ($v_{m,0}$ in Eq. \eqref{eq_Rudner_Ham}).


We now address the question of where such a 1D system of fermions may be found that has the bosonic description of Eq. \eqref{eq_dJC_RM}. Since the relative strength of the two parameters $\mu$ and $\lambda$ dictate the topological phase transition, we wish to find out what these two parameters represent in a 1D system. A famous model of interacting fermions in 1D is \cite{kopietz1997bosonization},
\begin{eqnarray}
\hat{H} = \hat{H}_0 + \hat{H}_{int},  \ \ \ \ \ \  \hat{H}_0 = \sum\limits_{k,\alpha} \epsilon_{k,\alpha} \hat{\psi}^{\dagger}_{k \alpha} \hat{\psi}_{k \alpha'}, \ \ \ \ \ \ \label{eq_H_0i}\\
 \hat{\rho}^{,\alpha}_{q} = \sum\limits_k \hat{\psi}^{\dagger}_{k \alpha} \hat{\psi}_{k +q, \alpha},  \ \ \ \ \ \ \ \ \ \ \ \ \ \ \ \ \ \ \label{eq_rho}\\
\hat{H}_{int}\{ \psi \} = \frac{1}{2V} \sum\limits_{q} \sum\limits_{\alpha \alpha'} f_q^{\alpha \alpha'} \hat{\rho}^{\alpha}_{-q} \hat{\rho}^{\alpha'}_{q}, \ \ \ \ \ \ \ \ \ \ \ \ \ \ \ \ \ \ \ \label{eq_Hint}\\
= \frac{1}{2V} \sum\limits_{k} \sum\limits_{q} \sum\limits_{\alpha \alpha'} f_q^{\alpha \alpha'} \hat{\psi}^{\dagger}_{k +q, \alpha} \hat{\psi}_{k \alpha} \rho^{\alpha'}_{q}, \ \ \ \ \ \label{eq_Hint_1}
\end{eqnarray}
where $\hat{\psi}_{k \sigma}(\hat{\psi}^{\dagger}_{k \sigma})$ is the canonical fermionic annihilation (creation) operator with wave-vector k and spin $\sigma$. We emphasize that there is no approximation involved in going from Eq. \eqref{eq_Hint} to Eq. \eqref{eq_Hint_1}, we are attempting to interpret an interacting fermionic system as a fermion-boson problem, (and later recast the problem into a spin-boson problem). This is related to a Hubbard-Stratonovich transformation \cite{yurkevich2002bosonisation,wetterich2007bosonic}. 
The non-interacting part may be written (`bosonized') in terms of boson operators $b_q$ from Eq. \eqref{eq_bq_dag} \cite{supp,giamarchi2003quantum},
\begin{eqnarray}
H_0 \simeq \sum_{q \neq 0} v_F|q|  b^{\dagger}_{q} b_{q}= \sum_{q \neq 0} H_q,\label{eq_Hbosons}
\end{eqnarray}
This is a famous result from 1D bosonization theory where the non-interacting bosonic dispersion is also linear, and have well defined energy and momentum for small $q$ \cite{supp,giamarchi2003quantum}. Quadratic fermionic interactions lead to free quadratic theory for the bosons, on par with interacting quartic fermionic interactions.

We may divide up in equal halves the Hamiltonian $H_q$ in Eq. \eqref{eq_Hbosons} to write it in a composite fermion-boson representation (we `partially' replace back the bosons i.e. only one of the two from Eq. \eqref{eq_bq_dag}),
\begin{eqnarray}
H_q = \frac{v_F|q| K(q) }{2} \sum_{k=-\infty}^{k_F} \left( b^{\dagger}_{q} c^{\dagger}_{k} c_{k+q} + c^{\dagger}_{k+q} c_{k} b_{q} \right), \ \ \ \ \ \label{eq_boson_fermi}
\end{eqnarray}
where Eq. \eqref{eq_b_sub} was used. Next, we employ Abrikosov's pseudo-fermion representation \cite{abrikosov1965electron,abrikosov1968formation,abrikosov1969magnetic,oppermann1973application} (functionally similar to the well-known Schwinger boson representation), 
\begin{eqnarray}
c^{\dagger}_{k+q}c_{k} = \sigma_q^+(k);\ \  {\rm and} \ \ \   
c^{\dagger}_{k}c_{k+q} = \sigma_q^-(k)
\label{eq_abrikosov_repr},
\end{eqnarray}
under the constraint, $ c^{\dagger}_{k+q} c_{k+q} + c^{\dagger}_{k} c_{k}=1 $. With a prefactor $J(q) = v_F |q| K(q)/2$ we arrive at,
\begin{eqnarray}
H_q = J(q) \sum_{k=-\infty}^{k_F} \left( b^{\dagger}_{q} \sigma_q^-(k) + \sigma_q^+(k) b_{q} \right)  \label{eq_JC_1D}
\end{eqnarray}
Writing $H_q=\sum_k H_{k,q}$ and again setting $q=1$ as in Eq. \eqref{eq_dJC_RM}, $H_k$ maps to the atom-cavity coupling term of the JC model with $J(q)\equiv\lambda$ in Eq. \eqref{eq_dJC_min}.

But, what of the drive on the atom -- the term $ \mu \sigma^+ + \mu \sigma^- $ in Eq. \eqref{eq_dJC_min}? We now show that interactions (added to the non-interacting model discussed so far) can lead to such drives on the envisioned spin/atom. We consider re-fermionization of the bosonization problem. As detailed in Ref. \cite{giamarchi2003quantum, supp}, in the theory of g-ology, four kinds of interactions are considered $g_1, g_2, g_3,$ and $g_4$, and two bosonic fields $\theta$ and $\phi$ are introduced. The $g_2$, $g_4$ interactions simply renormalize the coefficients in the quadratic bosonic Hamiltonian of the non-interacting model, $H = \frac{1}{2\pi} \int dx \left[ u K \left(\nabla \theta \right)^2 + \frac{u}{K} \left(\nabla \phi \right)^2 \right]$. The terms $u$ and $K$ are known in terms of $v_F$, $g_2$, and $g_4$ \cite{supp,giamarchi2003quantum}(this happens to be the set-up for Luttinger liquids, which are well-known to be found in 1D).

For example, the bosonic field $\phi_{\sigma}(x)$ is introduced from $g_{1\perp}$ interaction of the now-spinful fermions, $
H_{g_1} = g_{1\perp} \sum\limits_{\sigma} \left[\psi_{L,\sigma}^{\dagger} \psi_{R,\sigma} \psi_{R,-\sigma}^{\dagger} \psi_{L,-\sigma}  \right] 
= \frac{g_{1\perp}}{(2\pi\alpha)^2} \sum\limits_{s=\uparrow, \downarrow} \left[ e^{i\left(-2\phi_s(x) \right)} e^{i\left(2\phi_{-s}(x) \right)} \right]$, leading eventually to,
\begin{eqnarray}
H_{g_1} = \int dx \frac{2g_{1\perp}}{(2\pi\alpha)^2} \cos \left( 2\sqrt{2} \phi_{\sigma}(x)  \right) \ \ \ \ \ \ \label{eq_cos_phi_sig}
\end{eqnarray}
defined in terms of $\phi_{\sigma(\rho)}=\frac{1}{\sqrt{2}} \left[ \phi_{\uparrow}(x) \mp \phi_{\downarrow}(x) \right]$. The other field $\theta$ (commutes with $\phi$) is also introduced and the two pairs $(\phi_{\rho},\theta_{\rho})$ and $(\phi_{\sigma},\theta_{\sigma})$ obey the standard commutation relations.

The idea of refermionization hinges on interpreting the cosine terms as in Eq. \eqref{eq_cos_phi_sig} in terms of original fermionic operators. We consider the refermionzation of the Umklapp process $g_3$ (requires the presence of a lattice and periodic $\mathbf{k}$-space) that leads to the cosine interaction in terms of $\phi_{\rho}(x)$.
\begin{eqnarray}
H_b = \int dx \frac{2g_{3}}{(2\pi\alpha)^2} \cos \left( 2\sqrt{2} \phi_{\rho}(x) - \delta x  \right), \ \ \ \ \ \ \label{eq_cos_phi_rho}
\end{eqnarray}
where $\delta$ is the doping so that $4k_F = 2\pi/a + \delta$ with $q=\delta/2\pi$ leading to $e^{i4k_F x}=e^{i\delta x}$. The idea of refermionization is recognizing that
\begin{eqnarray}
\psi_R^{\dagger}(x)\psi_L(x)=\frac{1}{2\pi \alpha} e^{i2\phi(x)}
\end{eqnarray}
and a Hamiltonian may be found in terms of $c_{R,k}$ and $c_{L,k}$ \cite{supp,giamarchi2003quantum}. For our problem, we define a rescaled field $\phi'_{\rho}(x)$ such that $2\sqrt{2}\phi'_{\rho}(x)= 2\sqrt{2} \phi_{\rho}(x) - \delta x $. The corresponding fermionic operators would be,
$
\psi_k^{\dagger}(x)\psi_{k+q}(x)=\frac{1}{2\pi \alpha} e^{i2\phi(x)}
$
and they would contribute the term in the k-representation,
\begin{equation}
\pi \alpha \frac{2g_{3}}{(2\pi\alpha)^2} \sum_k \left( c^{\dagger}_{k+q} c_k + c^{\dagger}_{k} c_{k+q} \right) \label{eq_g3_mu}
\end{equation}
In the Abrikosov spin-representation (Eq. \eqref{eq_abrikosov_repr}), they contribute to the driven JC Hamiltonian in Eq. \eqref{eq_dJC_min}, the drive term $\mu \sigma^+ + \mu \sigma^-$ ($\mu$ being the prefactor in Eq. \eqref{eq_g3_mu}).
In summary, the non-interacting part in Eq. \eqref{eq_H_0i} leads to the JC term, and $g_1, g_3$ interactions lead to the atom driving part in the LJC model of Eq. \eqref{eq_dJC_min}. This accomplishes our first objective-- to show the 1D fermion-boson duality may bridge the gap between interacting fermionic systems and light-matter interaction type spin-boson model that possess CS eigenstates (see also \cite{biswas2024jaynes}).



\begin{figure*}[t]
\includegraphics[width=18cm]{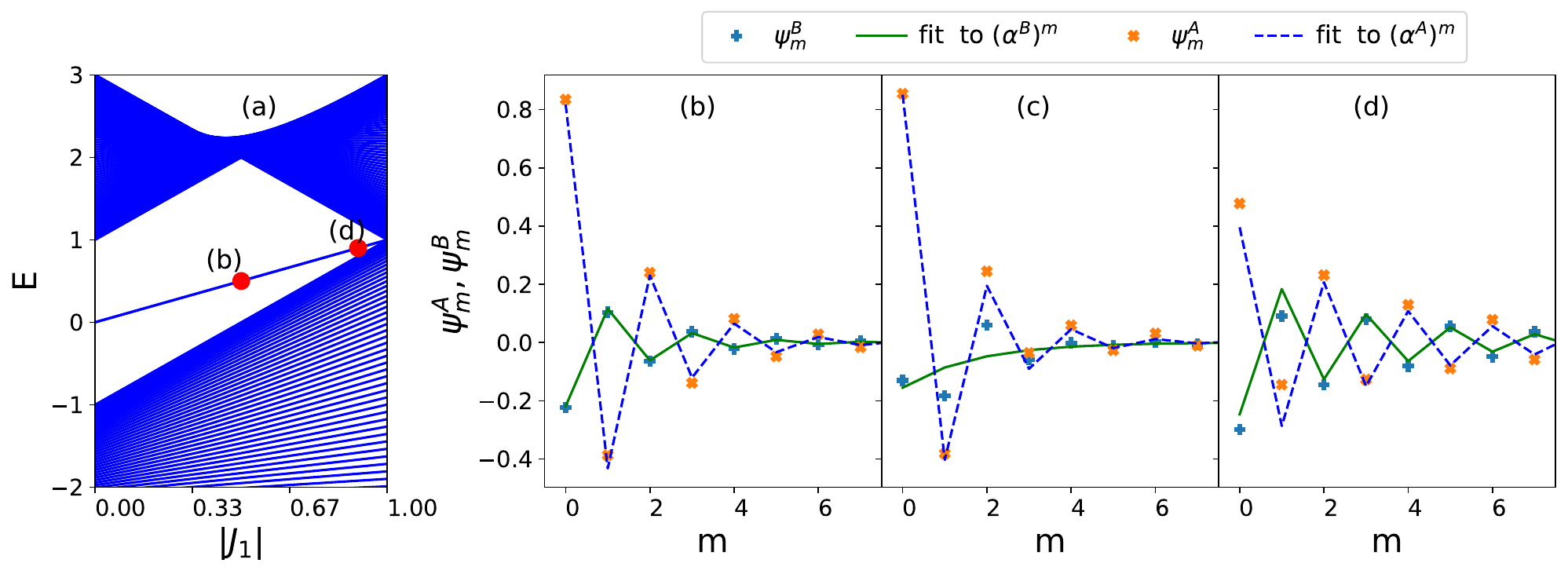}
\caption{\label{fig:1p} \textbf{(a)} Eigenspectrum of the fSSH model appended with $H_c^{(1)}$ with varying strength of $|J_1|$, $t_{\rm inter}=-2, t_{\rm intra}=-1$. The mid-gap ES energy linearly shifts as $E_{ES}=-2t_{\rm intra}/t_{\rm inter}J_1$ (Ref. \cite{longhi2018probing}) and merges with the bulk bands at $J_1=-1$. \textbf{(b,c,d)}: The ES wavefunction at (b) $J_1=-0.5, J_2=0$,  (c)$J_1=-0.5, J_2=-0.5$,  and (c) $J_1=-0.9, J_2=0$  calculated numerically (`+' and `x' markers). Lines represent best fits to Eq. \eqref{eq_left_edge_state} for each sublattice A and B. With $J_2=0$,  the fitting is highly accurate for small $|J_1|$ and becomes less accurate only when $E_{ES}$ approaches the bulk bands ($|J_1|\approx 1$), conforming to the unentangled product shape in Eq. \eqref{eq_product_state} throughout. Introduction of next-nearest-neighbor coupling $J_2$ in the SSH model, equivalent to two-photon drive in the mapped JC model, worsens the fitting much faster than $J_1$ as well as entangles the ES with the sublattice degree of freedom. More plots can be found in Figs. 6, 7 and especially Fig. 8. of the Supplemantary Material \cite{supp}.}
\end{figure*}

\section{Map using f-oscillators}
For the second approach, we do not map the bosonic operator $b$ to fermionic density fluctuations (Eq. \ref{eq_b_sub}) but instead look for a modification to the JC Hamiltonian itself, which directly hosts eigenstates that are edge states (ES) possessing a geometric series form of Eq. \ref{eq_left_edge_state}. Such an eigenstate has been reported in quantum optics literature as a nonlinear coherent state (NLCS) of the nonlinear Jaynes Cummings model (NLJC), \cite{larson2022jaynes, man1997f},
\begin{eqnarray}
\mathcal{H}^n_{JC} = \left(\lambda \sigma^+ a f(a,a^{\dagger}) + \lambda \sigma^- f(a,a^{\dagger}) a^{\dagger} \right) \ \ \  \label{eq_nl_dJC_min}
\end{eqnarray}
where we use bosonic mode $a$ instead of $b$ to distinguish from Eq. \eqref{eq_dJC_min}. With this intensity-dependent nonlinearity $f(a, a^\dagger)$, the Hamiltonian in Eq. \eqref{eq_dJC_min} (now generalized to NLJC) can also generate a SU(1,1) nonlinear coherent state (NLCS) \cite{miry2012generation,karimi2014quantum,supp}. These NLCS are defined with respect to a deformed displacement operator with the aid of the deformation function $f(\hat{n}=\hat{a}^{\dagger}\hat{a})$ \cite{man1997f,lopez2000photon,recamier2008nonlinear}.
They are also referred to as quantum mechanical f-oscillators, with deformed creation and annihilation operators \cite{man1997f},
\begin{eqnarray}
\hat{A}^{\dagger}=f(\hat{n})a^{\dagger},\ \ \hat{A}=\hat{a}f(\hat{n}). \ \ \ \ \label{eq_A_0}
\end{eqnarray}
Here we are interested in the choice of $f(\hat{n})=1/\sqrt{\hat{n}}$ (the ``Susskind-Glogower'' operator \cite{leon2011nonlinear,susskind1964quantum}) or the so-called quantum mechanical phase operator \cite{shapiro1991quantum}, 
\begin{equation}
A|n \rangle=|n-1\rangle,\ \ A^{\dagger}|n \rangle=|n+1\rangle, \ \ A|0 \rangle = 0   \label{eq_A}
\end{equation}
In addition, the coherent states become $|\alpha \rangle= N_1(|\alpha|^2)^{-1/2}\sum_{n=0}^{\infty} \alpha^n |n \rangle $, i.e. a ``Susskind-Glogower CS'' (\cite{leon2011nonlinear,gazeau2021generalized}) having the same geometric series form as Eq. \eqref{eq_left_edge_state}, that is, without the $\sqrt{n!}$ of a coherent state's Fock-basis expansion in the denominator. Therefore, the ES of the fSSH model may be thought of as a NLCS of a deformed oscillator. An injective mapping may be envisioned from LJC model to NLJC, where CS of the former maps to the NLCS of the latter. This mapping would be the same one (in a matrix form) as the mapping to a fSSH model discussed before i.e. ``the first mapping''. And, the $k$ indices in the model $\mathcal{H}^{(1)}_{JC}$ in Eq. \eqref{eq_dJC_RM} are in fact the photon number index $n$ in Eq. \eqref{eq_b}. In essence, the two mappings lead to the same matrix representation of the Hamiltonians in their respective bases (the $\{A_n, B_n\}$ site basis for the fSSH model, and the deformed Fock-state basis for the f-oscillators), although the constituent Hilbert spaces are different.  

For the second mapping, we now consider the question if such a mapping to a Fock-state basis (without the $\sqrt{n}$ coupling) actually provides us with a meaningful Hilbert space of basis states, which is required for quantum mechanical consistency. 
Fortunately, this result already exists in literature \cite{ali2004representations}. We elucidate it in the supplementary material \cite{supp} and provide a succinct summary here. Briefly, we can consider a set of nested Hilbert spaces, formally called a Gelfand triple: $\mathfrak{h}_F$, $\mathfrak{h}$, and $\mathfrak{h}_{F^{-1}}$ (with respective orthonormal basis sets $\{\phi^F_n\}, \{\phi_n\}, \{\phi^{F^{-1}}_n \}$) for giving NLCS a proper quantum mechanical standing. Each has standard oscillator operators $\{ a, a^{\dagger}, N \}$ and subscripted with $F^{-1}$ and $F$ that can be considered in other's Hilbert spaces as well. If $b$ in Eq. \eqref{eq_dJC_min} is assumed to be $a_F$ in the Hilbert space $\mathfrak{h}_F$, then in the Hilbert space $\mathfrak{h}$, it becomes the deformed annihilation operator $A$ \cite{supp,ali2004representations, man1997f},
with $A\phi_n = f(n)\sqrt{n} \phi_{n-1}, \ \ A^{\dagger}\phi_n = f(n+1)\sqrt{n+1} \phi_{n+1}$ (as in Eqs. \eqref{eq_A_0}, \eqref{eq_A} with $f(n)=1/\sqrt{n}$). In the Hilbert space $\mathfrak{h}$, $\eta_z^F= \mathcal{N}(|z|^2)^{-\frac{1}{2}} \sum\limits_{n=0}^{\infty} \frac{z^n \phi_n}{f(n)! \sqrt{n!}}$ \cite{ali2004representations,supp} is a NLCS, but in it's own Hilbert space $\mathfrak{h}_F$ it is a proper canonical CS with normalization well understood from quantum optics.

\section{Coherent drives and their effect on edge states}
Finally, we consider adding two extra drives to the linear JC model (Eq. \eqref{eq_Rudner_Ham}) and analyze their effects on the topological protection of SSH edge states and the immunity of CS to entanglement.
\begin{eqnarray}
H^{(1)}_{c} &=& i \left( Fb^{\dagger} - F^*b \right) \label{eq_1p}\\
H^{(2)}_{c} &=& i \left( Gb^2 - G^* {b^{\dagger}}^2 \right) \label{eq_2p}
\end{eqnarray}
$F$  and $G$ represent respectively a single-photon drive and a two-photon parametric drive, and they are classical coherent state amplitudes themselves. 
With the mapping defined by Eq. \eqref{eq_b_sub}, the translationally symmetric models will have $J_{\pm 1}= F$ from $H^{(1)}_{c}$ and $J_{\pm 2}= G$ from $H^{(2)}_{c}$. Hopping within the same sublattice breaks particle-hole symmetry \cite{ahmadi2020topological} as well as chiral symmetery, but not inversion symmetry \cite{longhi2018probing,jiao2021experimentally}. Both models can have an edge state in the topologically nontrivial regime; however, the energy of such an edge state is not pinned to zero (or the midgap) due to the breaking of inversion symmetry \cite{longhi2018probing}. We are interested in the question if they can have the geometric series shape of Eq. \eqref{eq_left_edge_state}, an ES of the form of a CS in the bosonic basis.

If the CS amplitude $F$ was replaced by a bosonic mode in $H^{(1)}_{c}$, it would become a beam-splitter Hamiltonian, and such transformations are not expected to get the mode $b$ entangled \cite{haroche2006exploring}. In fact, Ref. \cite{roos2008ion} showed that drives such as in $H^{(1)}_{c}$  do not corrupt the CS form even if their amplitude is time dependent. 

The nonlinear process in $H^{(2)}_{c}$, on the other hand, is well known to produce degenerate squeezing through a parametric process, and the Poissonian distribution of the CS in the original bosonic Fock-state lattice basis is lost \cite{teich1990squeezed, lvovsky_squeezed_2014, walls_squeezed_1983, dutt_-chip_2015, andersen_30_2016}. In the mapped translationally symmetric fSSH model, this corresponds to a loss of the geometric series form of the CS. Note that the coherent state is not an eigenstate of either $H^{(1)}_{c}$ or $H^{(2)}_{c}$. 

To numerically verify that these well-known quantum optical insights are preserved by the translationally invariant maps, we calculate the eigenstates and eigenvalues of an fSSH model with either  $H^{(1)}_{c}$ or  $H^{(2)}_{c}$ or both appended, while varying the strengths of $J_{\pm 1}=F$ and $J_{\pm 2}=G$. We fix the intercell and intracell hopping strengths of the unmodified inversion-symmetric fSSH model at $t_{\rm inter}=-2, t_{\rm intra}=-1$. 
Our calculations confirm that the eigenstate components $\psi_{A, B}^m$ can be fitted with Eq. \eqref{eq_left_edge_state} satisfactorily as long as $E_{ES}$ is sufficiently away from the bulk bands (Fig. \ref{fig:1p}b, $J_1=-0.5$). These regions are where a coherent state would be supported in the corresponding bosonic model. For larger value of $J_1$, the $E_{ES}$ approach the bulk bands and the quality of the fit to Eq. \eqref{eq_left_edge_state} is degraded (Fig. \ref{fig:1p}d, $J_1=-0.9$). The turning on of $J_2$ on the other hand makes this fitting to Eq. \eqref{eq_left_edge_state} unsatisfactory very quickly, even for small values of $J_2$ (Fig. \ref{fig:1p}c, $J_2=-0.5$). The $H^{(2)}_{c}$ model possesses the same symmetries as $H^{(1)}_{c}$, and has a topological edge state, but does not entail an ES that maps to a coherent state, that is, one that follows a geometric series form in the translationally invariant map \cite{supp}. More examples of the variation of the eigenstates for various values of $J_1$ and $J_2$ are shown in Figs. 6, 7 and especially Fig. 8 of the Supplementary material \cite{supp}. In the quantum trajectory \cite{gneiting_unraveling_2022, macri_revealing_2022} method of quantum dynamics, the system wavefunction evolves nonunitarily according to the effective non-Hermitian Hamiltonian, $H_{\rm eff}=H_c-i\frac{\gamma}{2}a^{\dagger}a$ between quantum jumps. In our fermionic model, this adds constant (imaginary) terms to the diagonal terms, which do not alter the eigenstates, or topological invariants and the CS eigenstate. Thus the requirements of exact pointer states under photon loss can be fulfilled trivially \cite{supp},\cite[chapter-4.4.5]{haroche2006exploring}. In the Supplemental material \cite{supp}, we prove that the evolution of the edge state in the fermionic model preserves the product structure,
\begin{eqnarray}
| \alpha \rangle(t) \otimes \begin{bmatrix} c_0\\c_1 \end{bmatrix} = | \alpha e^{-\left(i\omega_{\alpha} + \frac{\gamma}{2}  \right)t } \rangle  \otimes \begin{bmatrix} c_0\\ c_1 \end{bmatrix}, \ \ \ \label{eq_product_state}
\end{eqnarray}
which is in exact analogy to a cavity mode evolving under photon loss decay channel, $L_i = \sqrt{\gamma}a$. The $c_i$ coefficients are the support in the two sublattices B and A.  Moreover, we show that the product structure is not violated for nonzero $J_1$ even if large, which indicates the robustness of coherent states to entanglement with the environmental loss channel, but is violated for any  nonzero $J_2$ since the corresponding Fock-space lattice Hamiltonian has parametric two-photon drive or equivalently, squeezing terms. The breakdown of the product state form signifies entanglement with the environment as expected for most states other than coherent states, since the coherent state structure (or geometric series form in the mapped translationally invariant model) is not preserved in the presence of these squeezing interactions  \cite{supp}.


\section{Conclusions}

We have shown that there is a intricate connection between coherent states of electromagnetic fields and exponentially localized topological edge states. While we assumed optical photons for our analysis since the quantum electrodynamics (QED) of an atom-cavity coupled system is arguably the most relevant scenario, this connection should extend to any bosonic excitation by studying Glauber's coherent state forms \cite{glauber_coherent_1963} of the corresponding harmonic oscillator mode. Examples include microwave cavity photons in circuit QED, and resonator phonons coupled to superconducting qubits in quantum acousto-dynamics \cite{manenti_circuit_2017, lee_strong_2023, ruan_tunable_2024, banderier_unified_2023}, potentially including spins \cite{raniwala_piezoelectric_2023, joe_high_2024}, as long as the physical structure supports a mechanism of coherently driving the two-level atom-like system.
As a general rule, we found that the associated fermion model requires a midgap edge state. We observed that edge states obeying a geometric series form in the mapped fermionic model are equivalent to coherent states in the original bosonic model in a Fock-space lattice, and thus were observed to understandably resist entanglement with the environment. It is possible that future work will reveal other existing edge states of different shapes that could also resist entanglement with the environment, and our analysis leaves this question open. However, we establish, both through quantum optical arguments and through numerical calculations that a parametric two-photon drive or degenerate squeezing terms lead to rapid breakdown of the coherent state form, as reflected in the bosonic lattice and in the mapped fermionic model. Thus, the fermion-boson duality mapping we have harnessed in this paper also reveals that a parametric two-photon drive does not resist entanglement with the environment for an edge state. 
We anticipate that this fermion-boson duality can be extended to arrays of coupled cavities, with each cavity potentially supporting multiple bosonic modes \cite{banderier_unified_2023} or containing more than one atom to unravel further interconnections between quantum optics and topological physics.


\section*{Acknowledgments}

This work was supported by a National QLab joint seed grant from IonQ and the University of Maryland, a NAWCAD seed grant, and an NSF QuSeC-TAQS grant \# 2326792.

\appendix




\bibliography{coh_SSH, My_Library}

\end{document}